\documentclass[a4paper,11pt]{article}

\usepackage{named}

\newcommand{\szkew}[1]{\relax \setbox0=\hbox{\kern -24pt $\displaystyle#1$\kern 0pt }%
\box0}
{\catcode`\@=11 \global\let\ifjusthvtest@=\iffalse}

\newcounter{oldmycaption}

\newtheorem{theorem}{Theorem}[section]
\newtheorem{defined}{Definition}

\newtheorem{exa}[theorem]{Example}

\newtheorem{exe}{Exercise}

\newtheorem{pro}{Problem}
\newenvironment{problem}{\begin{pro} \rm }{\end{pro}}

\newcounter{symbol}
\newcommand{\indexsyma}[1]%
{\stepcounter{symbol}\index{zzz1 \thesymbol @\protect#1}}
\newcommand{\indexsymb}[1]%
{\stepcounter{symbol}\index{zzz2 \thesymbol @\protect#1}}
\newcommand{\indexsymc}[1]%
{\stepcounter{symbol}\index{zzz3 \thesymbol @\protect#1}}
\newcommand{\indexsymd}[1]%
{\stepcounter{symbol}\index{zzz4 \thesymbol @\protect#1}}
\newcommand{\indexsyme}[1]%
{\stepcounter{symbol}\index{zzz5 \thesymbol @\protect#1}}

\newcommand{\almazero}{{\sf Alma-0}}
\newcommand{\aaa}{{\sf AAA}}

\title{Programming in \almazero{}, or Imperative and Declarative Programming Reconciled}
\author{Krzysztof R. Apt\\
\normalsize {\em CWI } \\
\normalsize {\em P.O. Box 94079, 1090 GB Amsterdam, The Netherlands  } \\
\normalsize and \\
\normalsize {\em  Dept.\ of Mathematics, Computer Science, Physics \& Astronomy } \\
\normalsize {\em University of Amsterdam, The Netherlands  }\\[2mm]
Andrea Schaerf \\
\normalsize {\em Dipartimento di Ingegneria Elettrica, Gestionale e Meccanica} \\
\normalsize {\em Universit\`a di Udine} \\
\normalsize {\em via delle Scienze 208, 33100 Udine, Italy}
}
\date{}

\begin{document}

\maketitle

\begin{abstract}
  In \cite{ABPS98a} we introduced the imperative programming language
  \almazero{} that supports declarative programming. In this paper we
  illustrate the hybrid programming style of \almazero{} by means of
  various examples that complement those presented in \cite{ABPS98a}.
  The presented \almazero{} programs illustrate the versatility of the
  language and show that ``don't know'' nondeterminism can be
  naturally combined with assignment.
\end{abstract}

\section{Introduction}
\label{sec:intro}

Logic programming languages, notably Prolog, rely on two
important features: nondeterminism and unification.
The form of nondeterminism used is usually
called ``don't know'' nondeterminism.
According to it {\em some\/} path in the computation tree 
should lead to a correct outcome. 

There have been some efforts to incorporate
this form of nondeterminism into the imperative programming
paradigm. For early references see \cite{Coh79}. More recent
examples are the languages  Icon of  \cite{GG83}) 
and SETL of  \cite{SDDS86}. 

In \cite{ABPS98a} we pursued this approach to programming by
proposing another, simple, imperative language \almazero{} that
supports this form nondeterminism.  

Our rationale was that almost 25 years of experience with logic
programming led to an identification of the programming techniques
that make it a distinct programming paradigm. The imperative
programming constructs that support nondeterminism should support
these programming techniques in a natural way.

And indeed, we found that a number of logic programming jewels could be
reproduced in \almazero{} even though unification in the language is
limited to bare minimum and the language offers no support for
symbolic programming. 

But we also found that other programs, such as the solution to the
{\em Eight Queens\/} problem, could be coded in \almazero{} in a more
natural way than the logic programming paradigm permits.  Also, some
programs, such as the solution to the {\em Knapsack\/} problem, seem
to be very natural even though they use both nondeterminism and
assignment.

So the hybrid  programming style of \almazero{} apparently calls for new
programming techniques that need to be better understood and
explored. This is the aim of this paper that can be seen as a companion
article of \cite{ABPS98a}.

To this end we provide here a number of \almazero{}
programs that show versatility of the language and provide further
evidence that the constructs of the language encourage a natural style
of programming.
In particular, \almazero{} programs without assignment are declarative
in the sense that they admit a dual reading as a logic formula. 

It should be clarified that in general two types of nondeterminism 
have been considered in programming languages, ``don't know''
nondeterminism and ``don't care'' nondeterminism. According to the 
latter one {\em each\/} path in the computation tree should lead
to a correct outcome. This form of nondeterminism is present in
the guarded command language of \cite{Dij75}. It leads to
different issues and different considerations.

The paper is organized as follows. In Section~\ref{sec:alma-0} we
recall the basic elements of \almazero{}. In the remainder of the
paper we provide selected examples of \almazero{}
programs that complement those presented in \cite{ABPS98a} and
illustrate its use in different contexts.  More
specifically, in Section~\ref{sec:graph} we present two versions of a
classical graph traversal problem, namely the {\em longest path\/}
problem. In Section~\ref{sec:negation} we show how a typical feature
of the logic programming paradigm, namely {\em negation as failure},
can be also profitably exploited in \almazero{}.  Next, in
Section~\ref{sec:executable-specs} we illustrate how executable
specifications can be written in \almazero{}.  In
Section~\ref{sec:timetabling} we provide a more complex example of
\almazero{} programming by describing a solution to a classical
scheduling problem.  Finally, in
Section~\ref{sec:conclusions} we draw some conclusions and describe
the current status of the {\sf Alma} project.

\section{The language \almazero{}}
\label{sec:alma-0}

\almazero{} is an extension of a subset of Modula-2 that includes nine new
features inspired by the logic programming paradigm. We briefly recall
most of them here and refer to \cite{ABPS98a} for a detailed
presentation.

\begin{itemize}
\item Boolean expressions can be used as statements and vice versa.
  A boolean expression that is used as a statement and evaluates to
  {\tt FALSE} is identified with a {\em failure}.

\item {\em Choice points} can be created by the two nondeterministic
  statements {\tt ORELSE} and {\tt SOME}. The former is a dual of the
  statement composition and the latter is a dual of the {\tt FOR}
  statement.  Upon failure the control returns to the most recent
  choice point, possibly within a procedure body, and the computation
  resumes with the next branch in the state in which the previous
  branch was entered.
  
\item The created choice points can be erased or iterated over by
  means of the {\tt COMMIT} and {\tt FORALL} statements. {\tt COMMIT S
    END} removes the choice points created during a successful
  execution of {\tt S}.  {\tt FORALL S DO T END} iterates over all
  choice points created by {\tt S}. Each time {\tt S} succeeds, {\tt
    T} is executed.

\item The notion of {\em initialized} variable is introduced and the
  equality test is generalized to an assignment statement in case one
  side is an uninitialized variable and the other side an expression
  with known value.  The {\tt KNOWN} relation is introduced to test
  whether a variable of a simple type is initialized.

\item A new parameter passing mechanism, called {\em call by mixed
    form}, is introduced for variables of simple type. It works as
  follows: If the actual parameter is a variable, then it is passed by
  variable.  If the actual parameter is an expression that is not a
  variable, its value is computed and assigned to a new variable $v$
  (generated by the compiler): it is $v$ that is then passed by
  variable.  So in this case the call by mixed form boils down to call
  by value.

  This parameter mechanism, denoted by {\tt MIX}, is introduced to
  allow us to pass both values and uninitialized variables as actual
  parameters.
\end{itemize}

To clarify these extensions and \almazero{} programming style consider
the following problem from \cite{Gar79}.

\begin{problem}
Ten cells numbered $0,..,9$ inscribe a 10-digit number such that
each cell, say $i$,  indicates the total number of occurrences of
the digit $i$ in this number. Find this number.
\end{problem}

Here is a
simple solution to it in \almazero{}.

\newpage
\small
\begin{verbatim}
MODULE tendigit; 
VAR i, j, k, l, count, sum: INTEGER;
    a: ARRAY [0..9] OF INTEGER;
BEGIN
  FORALL
    sum := 0;
    FOR i := 0 TO 9 DO
      SOME j := 0 TO 10-sum DO
        a[i] = j;
        sum := sum + j
      END;
    END;
    sum = 10;
    FOR k := 0 TO 9 DO
      count := 0;
      FOR l := 0 TO 9 DO
        IF a[l] = k THEN count := count + 1; a[k] >= count END;
      END;
      a[k] = count
    END
  DO
  FOR i := 0 TO 9 DO WRITE(a[i]) END
  END
END tendigit.
\end{verbatim}
\normalsize

To better understand this program first note that any 10-digit number
that is a solution to this problem has the property that the sum of
its digits is 10.

Now, the first {\tt FOR} loop nondeterministically generates 10-digit
numbers, written as an array, with this property. This is done by
means of a {\tt SOME} statement.  The equality {\tt a[i] = j} is used
here as an assignment, while the equality {\tt sum = 10} is used as a
test.

The second {\tt FOR} loop tests whether a candidate array is a
possible solution. The testing can be abandoned if for some {\tt k}
the count exceeds the value {\tt a[k]}. This explains the use of the
test {\tt a[k] >= count}.

The above described code is within the {\tt FORALL} statement, so all
solutions to the problem are generated and each of them is printed.
The program yields the unique solution, namely 6210001000.

The still unexplained features of \almazero{} will be discussed later.

\section{Graph Traversal}
\label{sec:graph}

We now illustrate by means of two examples how \almazero{} can be used
in a natural way for graph-related problems. 

\subsection{Knight's Tour}
\label{subsec:knight-tour}

We begin with the following well-known problem.

\begin{problem}\label{pro:knight-tour}
Find a knight's tour on the $n\times n$ chess board in which each field
is visited is exactly once.
\end{problem}

Here is a solution in \almazero{}.

\begin{small}
\begin{verbatim}
MODULE KnightTour;
CONST   
  N = 5;
TYPE 
  [1..N] = [1..N];
  Board = ARRAY [1..N], [1..N] OF [1..N*N];

PROCEDURE Next(VAR row, col: INTEGER);
VAR i, j: INTEGER;
BEGIN
  EITHER i = 2;  j = 1
  ORELSE i = 1;  j = 2
  ORELSE i = -1; j = 2
  ORELSE i = -2; j = 1
  ORELSE i = -2; j = -1
  ORELSE i = -1; j = -2
  ORELSE i = 1; j = -2
  ORELSE i = 2; j = -1
  END;
  row := row + i;
  col := col + j;
  (1 <= row) AND (row <= N);
  (1 <= col) AND (col <= N)
END Next;

VAR i, j, k: INTEGER;
  x: Board;
BEGIN
  x[1,1] = 1;
  i = 1; j = 1;
  FOR k := 2 TO N*N DO
    Next(i,j);
    x[i,j] = k
  END;
  Print(x)
END KnightTour.
\end{verbatim}
\end{small}

Here the {\tt Next} procedure nondeterministically generates the
coordinates of the next field, given the current one.  This is done
now by means of an {\tt ORELSE} statement that explores all 
eight possibilities in turn.

After a call to {\tt Next}
the (implicitly) incremented value of {\tt k} is assigned to this new
field. Note that this assignment, {\tt a[i,j] = k}, is performed by
means of an equality. This is crucial, as it also prevents that a
field is visited again.  Indeed, if this is the case then {\tt a[i,j]}
has already a value and the equality fails. In this case the
backtracking takes place and the next, if any, candidate field is
generated.

\subsection{Longest Path}
\label{subsec:longest-path}

In the {\em Knight's tour\/} problem the $n\times n$
chess board can be viewed as a graph in which the squares are the
nodes and the possible knight moves are the arcs. In this way the
knight tour problem accounts to finding a simple path of maximal
length. The length of this path equals  $n^2$, the number of nodes.

Consider now a more general problem of finding the longest path in
an arbitrary directed graph.

\begin{problem} Given a directed graph $G=(V,E)$ and two
nodes $v_1,v_2\in V$ find the longest simple path that starts in $v_1$ and ends
in $v_2$.
\end{problem}

Recall that this decision problem is NP-complete (see
 \cite[problem ND29, page 213]{GJ79}).

We assume that the graph is represented by its adjacency matrix. We
also employ an array for marking the visited nodes and for storing the
current longest path. In what follows we use the following type
declarations.

\begin{small}
\begin{verbatim}
  Graph = ARRAY [1..N],[1..N] OF BOOLEAN;
  PathMark = ARRAY [1..N] OF INTEGER;
\end{verbatim}
\end{small}

The basic building block that we use for traversing the graph is the
following function \texttt{Successor} that upon backtracking generates all
successors of a given node. The function fails if the node has no
successor.

\begin{small}
\begin{verbatim}
PROCEDURE Successor(G: Graph; X: Node): Node;
VAR i: Node;
BEGIN
  SOME i := 1 TO N DO
    G[X,i]
  END;
  RETURN i
END Successor;
\end{verbatim}
\end{small}

The following procedure \texttt{LongestPath} consists of some
initializations followed by a \texttt{FORALL} loop that explores all
possible paths. Inside the \texttt{FORALL} loop, each path is
constructed by an inner loop that searches exhaustively for unvisited
successors until it gets to the requested final node.

In contrast to Problem~\ref{pro:knight-tour}, we do not know the
length of the longest path in advance. Therefore we use here a
\texttt{WHILE} statement rather than a \texttt{FOR} statement for
constructing the path. In addition, for each generated path
we need to check its length against the currently longest
one.

A node \texttt{X} is viewed as unvisited as long as \texttt{Path[X]
  = 0}. When \texttt{X} is visited, \texttt{Path[X]} gets the value
\texttt{k} which represents the position of \texttt{X}
in the path.

\begin{small}
\begin{verbatim}
PROCEDURE LongestPath(G: Graph; InitNode, FinalNode: Node): PathMark;
VAR k, max: INTEGER;
    i: Node;
    Path, LongPath: PathMark;
BEGIN
  FOR i := 1 TO N DO Path[i] := 0 END;
  i := InitNode;
  k := 0;
  max := 0;
  FORALL 
    WHILE (Path[i] = 0) AND (i <> FinalNode) DO
      k := k+1;
      Path[i] := k;
      i := Successor(G,i) (* generate a successor
                             nondeterministically *)
    END
  DO
    IF (i = FinalNode) AND (k > max)
    THEN max := k; LongPath := Path END
  END;
  RETURN LongPath
END LongestPath;
\end{verbatim}
\end{small}

The longest path is delivered by means of the return value of the
procedure.  If no path between {\tt InitNode} and {\tt FinalNode}
exists, then the variable {\tt LongPath} remains uninitialized, and
thus the value returned is also an uninitialized array, which can be
tested within the calling procedure by using the built-in procedure
{\tt KNOWN}.

\section{Use of Negation}
\label{sec:negation}

One of the important notions in logic programming is {\em negation by
  failure}. It is, in a nutshell, a meta-rule that allows us to
conclude a negation of a statement from the fact that it cannot be
proved (using the resolution method used in logic programming).
Negation by failure is a very useful concept that allows us to write
some remarkably concise Prolog programs.  Also, it supports
non-monotonic reasoning.  Actually, the negation by failure mechanism
provides a computational interpretation of the latter, a feature other
main approaches to non-monotonic reasoning lack.

Negation by failure is supported in \almazero{}, as well.  In fact, as
in logic programming, it is the mechanism used to evaluate negated
statements.  Consequently, we can use it in \almazero{} in the same
way as in logic programming and Prolog.

In \cite{ABPS98a} we already presented a number of programs that
used negation. Here we show an \almazero{} solution to the proverbial
{\em Tweety problem}, one of the classical benchmarks for
non-monotonic reasoning.  Let us recall it.

The problem is to reason in the presence of default assumptions.
In the natural language they are often expressed by means of the
qualification ``usually''. In what follows the ``usual'' situations
are identified with those which are not ``abnormal''.

We stipulate the following assumptions.
\begin{itemize}
\item The birds which are not abnormal fly
(i.e., birds usually fly).
\item The penguins are abnormal.
\item Penguins and eagles are birds.
\item Tweety is a penguin and Toto is an eagle.
\end{itemize}

The problem is to deduce which of these two birds flies. Here is a
solution in \almazero{}, where the code for \texttt{Print} is omitted.

\begin{small}
\begin{verbatim}
MODULE penguin;
TYPE Animal = (Tweety, Toto);

PROCEDURE penguin(MIX x: Animal);
BEGIN 
  x = Tweety
END penguin;

PROCEDURE eagle(MIX x: Animal);
BEGIN 
  x = Toto
END eagle;

PROCEDURE ab(MIX x: Animal);
BEGIN 
  penguin(x)
END ab;

PROCEDURE bird(MIX x: Animal);
BEGIN
  EITHER penguin(x) ORELSE eagle(x) END
END bird;

PROCEDURE fly(MIX x: Animal);
BEGIN
  bird(x);
  NOT ab(x)
END fly;

VAR x: Animal;
BEGIN
  FORALL fly(x)
  DO Print(x)
  END
END penguin.
\end{verbatim}
\end{small}

The use of the {\tt MIX} parameter mechanism allows us to use each 
procedure both for testing and for computing, as in Prolog.
In particular, the call {\tt fly(x)} yields to a nondeterministic
computing of the value of {\tt x} using {\tt bird(x)} and subsequent
testing of it using {\tt NOT ab(x)}.

It is instructive to compare this program with the more compact
Prolog program (see, e.g., \cite[page 303]{Apt97}):

\begin{small}
\begin{verbatim}
penguin(tweety). 
eagle(toto).
ab(X) :- penguin(X). 
bird(X) :- penguin(X).    
bird(X) :- eagle(X).    
fly(X) :- not ab(X), bird(X). 
\end{verbatim}
\end{small}

While logically both programs amount to equivalent formulas we see
that it is difficult to compete with Prolog's conciseness.

Other natural uses of negation in \almazero{} can be found in
some other programs in this article.

\section{Executable Specifications}
\label{sec:executable-specs}

The next example shows that in some circumstances
\almazero{} yields programs that are more intuitive than
those written in Prolog.

In general, specifications can and do serve many different purposes.
The issue whether specifications should be executable or not has been
for a long time a subject of a heated discussion, see, e.g.
\cite{Fuc92}.  
We do not wish to enter this discussion here but we show how
\almazero{} supports executable specifications in a very natural way.

As an example, consider the problem of finding the
lexicographically next permutation, discussed in \cite{Dij76}.

To specify this problem recall that by definition a sequence $out_1,
\dots, out_N$ is a permutation of $in_1, \dots, in_N$ if for some
function $\pi$ from $[1..N]$ onto itself we have
$$out_1, \dots, out_N = in_{\pi(1)}, \dots, in_{\pi(N)}.$$

This definition  directly translates into the following \almazero{}
program:

\small \begin{verbatim}
TYPE Sequence = ARRAY [1..N] OF INTEGER;

PROCEDURE Permutation(VAR in, out: Sequence);
VAR pi: Sequence;
    i, j: INTEGER;
BEGIN
  FOR i := 1 TO N DO 
    SOME j := 1 TO N DO
      pi[j] = i
    END
  END; (* pi is a function from 1..N onto itself and ...        *)
  FOR i := 1 TO N DO 
    out[i] = in[pi[i]] 
  END  (* out is obtained by applying pi to the indices of in *)
END Permutation;
\end{verbatim}
\normalsize

The procedure \texttt{Permutation} provides, upon backtracking, all
permutations of the given input sequence.

Next, we need to define the lexicographic ordering.  Let us recall the
definition: the sequence $a_1, \dots, a_N$ precedes lexicographically
the sequence
$b_1, \dots, b_N$ if some $i$ in the range $[1..N]$ exists such that
for all $j$ in the range $[1..i-1]$ we have $a_j = b_j$, and $a_i < b_i$.

In \almazero{} we write these specifications as follows:

\small \begin{verbatim}
PROCEDURE Lex(a,b: Sequence);
VAR i, j: INTEGER;
BEGIN
  SOME i := 1 TO N DO
    FOR j := 1 TO i-1 DO
      a[j] = b[j]
    END;
    a[i] < b[i]
  END
END Lex;
\end{verbatim}
\normalsize

Now $b$ is the lexicographically next permutation of $a$ if
\begin{itemize}
\item $b$ is a permutation of $a$,
\item $a$ precedes $b$ lexicographically,
\item no permutation exists that is lexicographically between $a$ and $b$.
\end{itemize}

This leads us to the following procedure {\tt Next} that uses an
auxiliary procedure {\tt Between}, which checks whether a
permutation exists  between \texttt{a} and \texttt{b}:

\small \begin{verbatim}
PROCEDURE Between(a,b: Sequence);
VAR c: Sequence;
BEGIN
  Permutation(a,c);
  Lex(a,c);
  Lex(c,b)
END Between;

PROCEDURE Next(VAR a, b: Sequence);
BEGIN
  Permutation(a,b);
  Lex(a,b);
  NOT Between(a,b)
END Next;
\end{verbatim}
\normalsize

This concludes the presentation of the program.  Note that it is fully
declarative and it does not use any assignment.  It is obviously
hopelessly inefficient, but still it could be used on the example
given in Dijkstra's book, to compute that {\tt 1 4 6 2 9 7 3 5 8} is
the lexicographically next permutation of {\tt 1 4 6 2 9 5 8 7 3}.

It is interesting to see that the above program is invertible in the
sense that it can be also used to specify and compute the
lexicographically previous permutation. In fact, we can use for this
purpose the same procedure {\tt Next} --- it just suffices to pass now
the given permutation as the second parameter of the procedure {\tt
  Next}.  For this purpose both parameters are passed by variable in
the procedures \texttt{Next} and \texttt{Permutation}.

In this way we can compute for instance that {\tt 1 4 6 2 9 5 8 3 7}
is the lexicographically previous permutation of {\tt 1 4 6 2 9 5 8 7
  3}.

\section{A Scheduling Application}
\label{sec:timetabling}

We now show how \almazero{} can be employed to solve scheduling
problems. In particular, we introduce a specific scheduling problem
known as the \emph{university course timetabling} problem
and discuss its solution in \almazero{}.

\subsection{Problem Definition}

The course timetabling problem consists in the weekly scheduling for all the
lectures of a set of university courses in a given set of classrooms, avoiding
the overlaps of lectures having common students.  We consider the basic 
problem (which is still NP-complete). Many variants of this problem have been
proposed in the literature. They involve more complex constraints and usually
consider an objective function to be minimized (see \cite{Scha99}).

\begin{problem}
There are $q$ courses $K_1,\dots,K_q$, and each course $K_i$ consists of
$k_i$ required lectures, and $p$ periods $1..p$. For all $i\in 1..q$, all lectures
$l\in 1..k_i$ must be assigned to a period $k$ in such a way that the
following constraints are satisfied:

\begin{description}
\item[Conflicts:] There are $c$ curricula $S_1,\dots,S_c$, which are groups
  of courses that have common students. Lectures of courses in $S_l$ must be
  all scheduled at different times, for each $l\in 1..c$.
\item[Availabilities:] There is an availability binary matrix $A$ of size
  $q\times p$. If $a_{ij} = 1$ then lectures of course $i$ cannot be
  scheduled at period $j$.
\item[Rooms:] There are $r$ rooms available. At most $r$ lectures can be
  scheduled at period $k$, for each $k\in 1..p$. 

\end{description}
\end{problem}

\subsection{A solution in \almazero{}}

We now provide a solution of this problem in \almazero{}. We start
with the constant and type definitions necessary for the
program.

\begin{small}
\begin{verbatim}
CONST 
   Courses = 10;  (* p *)
   Periods = 20;  (* q *)
   Rooms = 3;     (* r *)
TYPE    
   AvailabilityMatrix = ARRAY [1..Courses],[1..Periods] OF BOOLEAN;
   ConflictMatrix = ARRAY [1..Courses],[1..Courses] OF BOOLEAN;
   RequirementVector = ARRAY [1..Courses] OF INTEGER;
   TimetableMatrix = ARRAY [1..Courses],[1..Periods] OF BOOLEAN;
\end{verbatim}
\end{small}

Conflicts are represented by a $q\times q$ matrix of the
type {\tt ConflictMatrix} such that the
element $(i,j)$ of the matrix is \emph{true} if courses $K_i$ and
$K_j$ belong simultaneously to at least one curriculum.

The solution is returned by means of a $q\times p$ boolean matrix of
the type \texttt{TimetableMatrix}. Each element $(i,j)$ of the matrix
is {\em true} if a lecture for the course $K_i$ is given at period
$j$ and \emph{false} otherwise.

The procedure \texttt{Timetabling} provides the solution of this problem
in \almazero{}.  It follows faithfully the specification of the
problem and it performs an exhaustive backtracking search for a
feasible solution.

For each course $K_i$ the procedure looks for a number of periods
equal to the number of lectures $k_i$ of the course.  The array
\texttt{BusyRooms} counts the number of rooms already used for each
period, and is used to check the room occupation constraints.

In order to avoid exploring symmetric solutions for the lectures of a
course, each lecture is always scheduled later than the previously
scheduled lectures of the same course. This is done by using the variable
\texttt{PeriodOfPre\-viousLecture} which keeps track of the period of the most
recently scheduled lecture.

\begin{small}
\begin{verbatim}
PROCEDURE Timetabling(Available: AvailabilityMatrix;
                      Conflict: ConflictMatrix;
                      Requirements: RequirementVector;
                      VAR Timetable: TimetableMatrix);
VAR
   BusyRooms : ARRAY [1..Periods] OF INTEGER;
   C, C1, L, P : INTEGER;
   PeriodOfPreviousLecture : INTEGER;
BEGIN
   FOR P := 1 TO Periods DO
      BusyRooms[P] := 0;
   END;
   FOR C := 1 TO Courses DO
      PeriodOfPreviousLecture := 0;
      FOR L := 1 TO Requirements[C] DO
         SOME P := PeriodOfPreviousLecture+1 TO Periods DO
            Available[C,P];
            BusyRooms[P] < Rooms;
            FOR C1 := 1 TO C-1 DO 
               NOT (Conflict[C1,C] AND Timetable[C1,P])
            END;
            Timetable[C,P] := TRUE;
            BusyRooms[P] := BusyRooms[P] + 1;
            PeriodOfPreviousLecture := P;
         END
      END
   END
END Timetabling;
\end{verbatim}
\end{small}

The proposed procedure can solve only relatively small instances of
the problem. For larger ones more complex algorithms and heuristic
procedures are needed (see \cite{Scha99}).

\subsection{Additional Functionalities}

If no solution to the given problem instance exists, it is in general
necessary to relax some of the constraints.  The following procedure
checks whether a solution exists when one single conflict
constraint is relaxed. If the solution of the relaxed instance of the
problem is found, its solution is returned along with the constraint
which has been relaxed. This constraint is returned by means of
two courses \texttt{c1} and \texttt{c2} which are no more considered
in conflict.

\begin{small}
\begin{verbatim}
PROCEDURE RelaxedTimetabling(Available: AvailabilityMatrix;
                             VAR Conflict: ConflictMatrix;
                             Requirements: RequirementVector;
                             VAR Timetable: TimetableMatrix;
                             MIX c1, c2: INTEGER);
VAR
   i, j: INTEGER;
BEGIN
   EITHER
     Timetabling(Available, Conflict, Requirements, Timetable)
   ORELSE
     SOME i := 1 TO Courses-1 DO
       SOME j := i+1 TO Courses DO
         Conflict[i,j];
         c1 = i; c2 = j;
         Conflict[i,j] := FALSE;
         Timetabling(Available, Conflict, Requirements, Timetable)
       END
     END
   END
END RelaxedTimetabling;
\end{verbatim}
\end{small}

Finally, the following procedure produces all relaxed and non-relaxed solutions
of the problem. The simple code for the procedures \texttt{Initialize}
and \texttt{PrintSolution} is omitted.

\begin{small}
\begin{verbatim}
PROCEDURE CreateTimetable;
VAR 
  Available: AvailabilityMatrix;
  Conflict: ConflictMatrix;
  Requirements: RequirementVector;
  Timetable: TimetableMatrix;
  NbrSolutions: INTEGER;
  c1, c2: INTEGER;
BEGIN
  Initialize(Available,Conflict,Requirements,Timetable);
  NbrSolutions := 0;
  FORALL
    RelaxedTimetabling(Available,Conflict,Requirements,Timetable,c1,c2)
  DO
    NbrSolutions := NbrSolutions + 1;
    WRITELN('Solution number ',NbrSolutions);
    PrintSolution(Available,Timetable);
    IF KNOWN(c1)
    THEN WRITELN('Conflict between course ', c1,' and ',c2,' relaxed')
    ELSE WRITELN('No constraint relaxed for this solution');
    END
  END;
  IF NbrSolutions > 0
  THEN WRITELN('Number of solutions : ',NbrSolutions)
  ELSE WRITELN('No solution found.');
  END;
  WRITELN
END CreateTimetable;
\end{verbatim}
\end{small}

Note the use of the built-in procedure \texttt{KNOWN} that checks
whether the variable \texttt{c1} is initialized or not. This test
allows us to check whether a constraint has been relaxed.

Finally, note that \texttt{c1} and \texttt{c2} are passed by
\texttt{MIX}. This way, not only a variable but also a constant can be
supplied as an actual parameter. For example, the following call searches
for a solution in which the possible relaxation involves course $K_1$:
\begin{small}
\begin{verbatim}
   RelaxedTimetabling(Available,Conflict,Requirements,Timetable,1,c);
\end{verbatim}
\end{small}

\noindent Here \texttt{c} is an uninitialized variable.
       
\section{Conclusions}
\label{sec:conclusions}

In this paper we presented a number of programs written in
\almazero{}.  They were chosen with the purpose of illustrating the
versatility of the resulting programming style.  The solution to some
other classical problems, such as $\alpha$-$\beta$ search, STRIPS
planning, knapsack, and Eight Queens, have been already provided in
\cite{ABPS98a}.

These programs show that imperative and logic programming can be
combined in a natural and effective way. The resulting programs are in
most cases shorter and more readable than their counterparts
written in imperative or logic programming style.

Let us review now the work carried out on \almazero{}.
The implementation of the language \almazero{} is based on an abstract
machine, called \aaa{}, that combines the features of a RISC
architecture and the WAM abstract machine.  In the current version the
\aaa{} instructions are translated into C code. The implementation is
described in~\cite{ABPS98a} and explained in full detail
in~\cite{Par97}.  The \almazero{} compiler is available via the Web at
\verb+http://www.cwi.nl/alma+.

An executable operational specification of a large fragment of
\almazero{} is provided using the ASF+SDF Meta-Environment of
\cite{Kli93}. This is described in \cite{ABPS98a}
and comprehensively explained in \cite{Bru98}.

An extension of \almazero{} that integrates constraints into the
language is the subject of an ongoing research.  Various issues related
to such integration are highlighted in \cite{AS99a}. In particular, the
role of logical and customary variables, the interaction between the
program and the constraint store, the local and global unknowns, and
the parameter passing mechanisms are considered there.

Finally, in \cite{AB99} a computational interpretation of first-order
logic based on a constructive interpretation of satisfiability w.r.t.
a fixed but arbitrary interpretation is studied. This work provides
logical underpinnings for a fragment of \almazero{} that does not
include assignment and allows us to reason about
\almazero{} programs written in this fragment.


\end{document}